\documentclass[preprint2]{aastex}

\shorttitle{The $^{15}$O($\alpha$,$\gamma$)$^{19}\textrm{Ne}$ reaction and the stability of thermonuclear burning}
\shortauthors{Fisker, Tan, G{\"o}rres, Wiescher, and Cooper}
\begin{document}
\title{The $^{15}$O($\alpha$,$\gamma$)$^{19}\textrm{Ne}$ reaction rate and the stability of thermonuclear burning on accreting neutron stars}
\author{Jacob Lund Fisker, Wanpeng Tan, Joachim G\"orres, and Michael Wiescher}
\affil{Department of Physics and Joint Institute for Nuclear Astrophysics, \\ University of Notre Dame, Notre Dame, IN 46556}
\email{jfisker@nd.edu, wtan@nd.edu, jgoerres@nd.edu, mwiesche@nd.edu}
\author{Randall L. Cooper\altaffilmark{1,2}}
\affil{Harvard-Smithsonian Center for Astrophysics, 60 Garden Street, Cambridge, MA 02138}
\email{rcooper@cfa.harvard.edu}
\altaffiltext{1}{Kavli Institute for Theoretical Physics, Kohn Hall, University of California, Santa Barbara, CA 93106}
\altaffiltext{2}{KITP Graduate Fellow}

\begin{abstract}
Neutron stars in close binary star systems often accrete matter from their companion stars. 
Thermonuclear ignition of the accreted material in the atmosphere of the neutron star leads to a thermonuclear explosion which is observed as an X-ray burst occurring periodically between hours and days depending on the accretion rate. 
The ignition conditions are characterized by a sensitive interplay between the accretion rate of the fuel supply and its depletion rate by nuclear burning in the hot CNO cycle and the $rp$-process.
For accretion rates close to stable burning the burst ignition therefore depends critically on the hot CNO breakout reaction ${}^{15}\textrm{O}(\alpha,\gamma){}^{19}\textrm{Ne}$ that regulates the flow between the hot CNO cycle and the rapid proton capture process. 
Until recently, the ${}^{15}\textrm{O}(\alpha,\gamma){}^{19}\textrm{Ne}$ reaction rate was not known experimentally and the theoretical estimates carried significant uncertainties. 
In this paper we perform a parameter study of the uncertainty of this reaction rate and determine the astrophysical consequences of the first measurement of this reaction rate. Our results corroborate earlier predictions and show that theoretically burning remains unstable up to accretion rates near the Eddington limit, in contrast to astronomical observations.
\end{abstract}

\keywords{X-rays: bursts --- nuclear reactions --- stars: neutron}

\section{Introduction}\label{sec:introduction}
Recently, many groups \citep{Woosley04,Amthor06,Cooper06, Fisker06,Fisker07a,Roberts06} have considered the nuclear reactions of the explosive thermonuclear runaway that lead to a type I X-ray burst on the surface of an accreting neutron star \citep[for reviews, see][]{Bildsten98c,Strohmayer06}. 
The ${}^{15}\textrm{O}(\alpha,\gamma){}^{19}\textrm{Ne}$-reaction is particularly interesting because it is the gateway between the hot CNO cycles, which are associated with stable burning during the quiescent phase, and the $rp$-process, which is associated with the thermonuclear runaway \citep{Wallace81}. 

When hydrogen and helium accrete onto the neutron star and advect into the atmosphere, 
hydrogen burns via the hot CNO cycle $^{12}\textrm{C}(p,\gamma)$ ${}^{13}\textrm{N}(p,\gamma)$ ${}^{14}\textrm{O}(\beta^+\nu_e)$ ${}^{14}\textrm{N}(p,\gamma)$ ${}^{15}\textrm{O}(\beta^+\nu_e)$ ${}^{14}\textrm{N}(p,\alpha)$ ${}^{12}\textrm{C}$. 
The hot CNO cycle is temperature independent ($\beta$-limited) and therefore its rate 
depends only on the abundance of ${}^{14}\textrm{O}$ and ${}^{15}\textrm{O}$.
Using a time-dependent X-ray burst model, \cite{Fisker06} investigated the importance of the ${}^{15}\textrm{O}(\alpha,\gamma)$ ${}^{19}\textrm{Ne}$ breakout reaction rate and showed that a lower breakout rate restricts the outflow from the hot CNO cycle so that the hot CNO cycle processes hydrogen to helium at a faster rate. 
This increases the energy output and thereby raises the temperature of the entire envelope so that ${}^{4}\textrm{He}$ burns to ${}^{12}\textrm{C}$ further out in the atmosphere, which in turn increases the efficacy of the hot CNO cycle.
This moves the hydrogen burning front (defined by $X_H=0$) further out where ${}^{4}\textrm{He}$ burns slower. The energy generation rate therefore decreases until the hydrogen burning front advects down to a sufficient depth and restarts this cycle.
 For a global accreting rate of $\dot{M}=1\times 10^{17}\,\textrm{g}\,\textrm{s}^{-1}$, this results in the non-bursting oscillatory luminosity found in time-dependent model simulations by \cite{Fisker06} 
 and later in the two-zone model of \citep{Cooper06b}. The result of this burning behavior is a copious production of ${}^{12}\textrm{C}$ \citep{Fisker06} which theoretically would provide the required fuel concentration to explain superbursts \citep{Cumming01}.
Unfortunately for this theory, astronomers frequently observe X-ray bursts at this accretion rate \citep[e.g.,][]{Cornelisse03b,Galloway06}, which suggests that the aforementioned burning cycle does not occur.
This leads to the astrophysically based conclusion that the  true $^{15}\textrm{O}(\alpha,\gamma){}^{19}\textrm{Ne}$ reaction rate must be higher than the lower limit calculated by \cite{Fisker06}.

In the first part of this paper, we therefore perform a parameter study of the $^{15}\textrm{O}(\alpha,\gamma){}^{19}\textrm{Ne}$ reaction rate as a function of the accretion rate to investigate the impact of uncertainty of the reaction rate. The results of this study corroborate earlier estimates of the critical accretion rate by \cite{Fushiki87,Bildsten98c,Fisker03,Woosley04,Heger05,Cooper06}. It also strengthens the astrophysically based conclusion of \cite{Fisker06} that set a lower limit on the ${}^{15}\textrm{O}(\alpha,\gamma)$ ${}^{19}\textrm{Ne}$ breakout reaction rate

In the second part, we use the recently measured rate of \cite{Tan07}, which includes an experimentally determined lower limit and repeat and improve the calculations of \cite{Fisker06}. We also take advantage of the significantly reduced uncertainty to determine the uncertainty range in the transition accretion rate between stable and unstable burning. 

We discuss the computational model in \S \ref{sec:model}. The parameter study is described in \S \ref{sec:parameterstudy} which is followed by a discussion of the implication of \cite{Tan07}'s new rate in \S \ref{sec:newrate} and the conclusion in \S \ref{sec:conclusion}.

\section{Computational Model}\label{sec:model}
The impact of the $^{15}$O($\alpha$,$\gamma$)$^{19}\textrm{Ne}$ rate has been investigated in the framework of a dynamical and self-consistent spherically symmetric X-ray burst model that has been used also in \cite{Fisker05b,Fisker05a,Fisker06,Fisker07a}. 
This model couples a modified version of a general relativistic hydrodynamics code~\citep{Liebendoerfer02} with a generic nuclear reaction network~\citep{Hix99} using the operator-split method~\citep{Henyey59}. 
The nuclear reaction flow and the conductive, radiative, and convective heat transport are 
computed in a general relativistic spherically symmetric geometry.

The reaction network comprises the same 298 isotope network between the valley of stability and the proton drip line as \cite{Fisker06}. 
Except for the $^{15}$O($\alpha$,$\gamma$)$^{19}\textrm{Ne}$-rate, all the proton-, neutron-, and alpha-induced reactions are adopted from the latest version of the REACLIB library~\citep{Sakharuk06} which was also used in \cite{Weinberg06}.

The radiative, conductive, and convective heat transport is treated in the formalism of \cite{Thorne77}. 
The radiative opacities due to Thompson scattering and free-free absorption are calculated according to \cite{Schatz99} and the conductivities for electron scattering on electrons, ions, phonons, and impurities are based on the work of \cite{Brown00}. 
The accreted matter is assumed to be fully ionized on impact. 
We use an arbitrarily relativistic and arbitrarily degenerate equation of state to describe the electron gas. 
Due to the shorter quantum wavelength of the nucleons, the nucleon gas behaves as an ideal gas.

The model code tracks energy transport with high precision and takes
into account the heat transport between the atmosphere and the neutron star
core~\citep{Brown04}. The thermal energy released from the core is $0.11\textrm{MeV}\,\textrm{nuc}^{-1}$ for $\dot{M}=1.12\times 10^{17}\textrm{g}\,\textrm{s}^{-1}$, but it generally depends on the accretion rate, the composition of the X-ray burst ashes, the temperature of the envelope, and the mass and radius of the core.
The neutron star core boundary corresponds to a pressure of $P = 7 \times 10^{23}\,\textrm{dyn}\,\textrm{cm}^{-2}$ and the core itself is characterized by a mass of $1.4M_\odot$ and a radius of $11.06\,\textrm{km}$ leading to a redshift of $1+z=(1-2GM/Rc^2)^{-1/2}=1.27$ and a gravitational acceleration of $g=(1+z)GM/R^2=1.8\times 10^{14}\textrm{cm}\,\textrm{s}^{-2}$.
Similarly, the upper atmosphere is described by a relativistically corrected grey atmosphere~\citep{Thorne77,Cox04} using a 4th order Runge-Kutta method to numerically integrate the hydrostatic heat and pressure equations from the model boundary out to $P = 10^{18}\,\textrm{dyn}\,\textrm{cm}^{-2}$.

The computation followed a sequence of accretion and burst phases until a limit cycle equilibrium was reached. 
This was performed for different $^{15}$O($\alpha$,$\gamma$)$^{19}\textrm{Ne}$ rates for a range of accretion rates as described in the next section.

\section{Parameter study of  ${}^{15}\textrm{O}(\alpha,\gamma){}^{19}\textrm{Ne}$ }\label{sec:parameterstudy}
\begin{figure}[tbph]\center\includegraphics[width=1.0\linewidth]{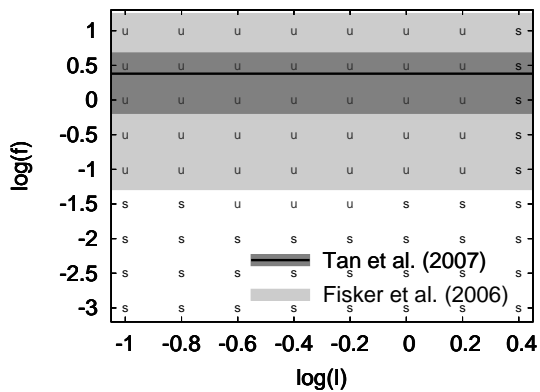}
\caption{The figure shows a log-log plot of the $^{15}$O($\alpha$,$\gamma$)$^{19}\textrm{Ne}$ rate of \cite{Caughlan88} scaled with a parameter, $f$, and the accretion rate scaled linearly with a parameter, $I$, so $\dot{M}=I\times 1.12\times 10^{18}\textrm{g}\,\textrm{s}^{-1}$. Each letter represents a simulation with \emph{u} showing bursting behavior and \emph{s} showing non-bursting behavior. For reference, the solid line shows the rate of the new measurement for $T=4\times 10^8\textrm{K}$. The light and dark shaded areas show the uncertainty at $T=4\times 10^8\textrm{K}$ of the estimate of \cite{Fisker06} and \cite{Tan07} respectively.}\label{fig:constraint}\end{figure}

Increased availability of computing power has significantly expanded our ability to perform parameter studies of full one-dimensional X-ray burst simulations. 
Fig.~\ref{fig:constraint} shows the results of a parameter study of 72 simulations where the $^{15}$O($\alpha$,$\gamma$)$^{19}\textrm{Ne}$ rate of \cite{Caughlan88}, which is based on the work of \citet{Langanke86}, has been scaled linearly with a parameter, $f$. 
Similarly the accretion rate has been scaled linearly with a parameter, $I$, where $I=1$ corresponds to the Eddington limit $\dot{M}_{\textrm{Edd.}}=1.12\times 10^{18}\textrm{g}\,\textrm{s}^{-1}$. The local Eddington limit is $\dot{m}_{\textrm{Edd.}}=\dot{M}_{\textrm{Edd.}}/(4\pi R^2)=6.62\times 10^{4}\textrm{g}\,\textrm{cm}^{-2}\,\textrm{s}^{-1}$ assuming that accretion is spherically symmetric.
The simulations were computed on an $8\times9$ grid corresponding to the \emph{u} (bursting behavior) and \emph{s} (non-bursting behavior) letters in Fig.~\ref{fig:constraint}. The grid covers the range between H/He ignited bursts and stable burning. In the following subsections we describe the different outcomes of this study.
\subsection{High values of $f$}\label{subsec:highfanyI}
For $\log(f)\gtrsim-1.5$ burning is unstable for accretion rates up to $\log(I)=$0.2--0.4 (see Fig.~\ref{fig:constraint}) and a H/He triggered burst occurs after the ${}^{15}\textrm{O}(\alpha,\gamma)$ ${}^{19}\textrm{Ne}(\beta^+,\nu)$ ${}^{19}\textrm{F}(p,\alpha)$ ${}^{16}\textrm{O}(p,\gamma)$  ${}^{17}\textrm{F}(p,\gamma)$ ${}^{18}\textrm{Ne}(\beta^+,\nu)$ ${}^{18}\textrm{F}(p,\alpha){}^{15}\textrm{O}$ cycle \citep[for details, see][]{Cooper06,Fisker07a} has been active for several hundreds of seconds before the runaway. Once the $3\alpha$ reaction triggers, the ${}^{19}\textrm{Ne}(p,\gamma)$ ${}^{20}\textrm{Na}$ breakout reaction starts the $rp$-process. This activates the ${}^{14}\textrm{O}(\alpha,p)$ ${}^{17}\textrm{F}$-reaction and creates a direct reaction flow from helium into the $rp$-process viz.~from $3\alpha$ to ${}^{12}\textrm{C}(p,\gamma)$ ${}^{13}\textrm{N}(p,\gamma)$ ${}^{14}\textrm{O}(\alpha,p)$ ${}^{17}\textrm{F}(p,\gamma)$ ${}^{18}\textrm{Ne}(\alpha,p)$ ${}^{21}\textrm{Na}$. This reaction circumvents the $T_{1/2}=76.4\,\textrm{s}$ $\beta^+$-decay of ${}^{14}\textrm{O}$ and allows a flow directly into the $rp$-process. 

For $\log(f)\geq -1.0$ burning becomes stable for $\log(I)=$0.2--0.4 independent of the exact value of the ${}^{15}\textrm{O}(\alpha,\gamma)$ ${}^{19}\textrm{Ne}$ reaction. This means that previous theoretical and computational estimates of the critical accretion rate \citep{Fushiki87,Bildsten98c,Fisker03,Woosley04,Heger05} which were based on the rate of \cite{Caughlan88}, i.e. $\log(f)\equiv0$, 
are independent of the exact value of the reaction rate. For $\log{f}\sim -1.5$, small uncertainties the ${}^{15}\textrm{O}(\alpha,\gamma)$ ${}^{19}\textrm{Ne}$ reaction rate would be sufficient to change the critical accretion rate by almost an order of magnitude \citep{Cooper06,Fisker06}.

\subsection{Low values of $f$ at low accretion rates}
For $\log(f)\lesssim-2$ and $\log(I)\lesssim -0.5$, the ${}^{15}\textrm{O}(\alpha,\gamma)$ ${}^{19}\textrm{Ne}$-reaction is too weak, so the above mentioned cycle never activates the ${}^{14}\textrm{O}(\alpha,p)$ ${}^{17}\textrm{F}$-reaction; the small amount of unstable helium fizzles, and the thermonuclear runaway never happens. This results in an oscillating behavior \citep{Fisker06}.

\cite{Cooper06} speculated that a low value of $f$ could stabilize bursts at lower accretion rates. Using the time-dependent model at low values of $f$, we find the same stable/unstable transition accretion rate for the helium trigger at $\log(I)\sim -0.5$ thus complimenting the calculations of \cite{Cooper06}. However, It should be noted that for $\log(f)\lesssim-2$ the instability of the helium trigger never results in a burst as explained in the preceding paragraph and the next subsection. Therefore this behavior is marked with an $s$ in Fig.~\ref{fig:constraint}. 

\subsection{Low values of $f$ at high accretion rates}
\begin{figure}[tbph]\center\includegraphics[width=1.0\linewidth]{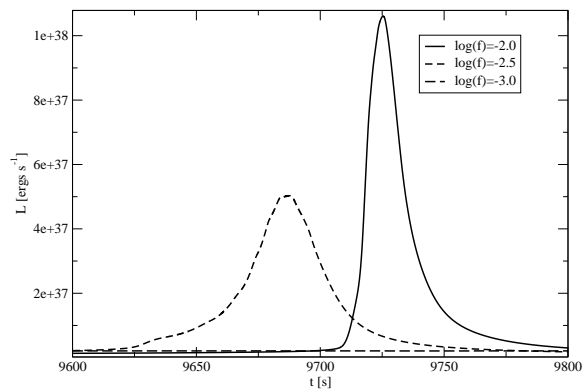}
\caption{This figure shows the luminosity for three different values of $f$ as a function of time for simulations with an accretion rate of $\log(I)=-0.4$. Notice the very long rise times and the symmetrical decays.}\label{fig:lowfhighmdot}\end{figure}

For $\log(f)\lesssim-2$ and $\log(I)\gtrsim -0.5$, the higher accretion rate increases compressional heating which heats the atmosphere to several hundred million degrees. Despite the low value of $f$ the natural temperature dependence of the ${}^{15}\textrm{O}(\alpha,\gamma)$ ${}^{19}\textrm{Ne}$-reaction means that helium burning triggers the $rp$-process. However, since the value of $f$ is small, the $rp$-process is fed very slowly resulting in increasingly longer luminosity rise times on the order tens of seconds as $f$ decreases with burning becoming stable at very low values of $f$ as shown in Fig.~\ref{fig:lowfhighmdot}. For these values of $f$, the accretion rate at which unstable burning transitions to stable burning is still uncertain.

Fig.~\ref{fig:lowfhighmdot} shows luminosity peaks with an almost symmetrical decay suggesting that the atmosphere is driven by an oscillating burning front rather than a thermonuclear burst followed by a thermal decay.
Such long rise times have never been observed at any accretion rate, so this result strengthens the argument for a lower limit on the $^{15}$O($\alpha$,$\gamma$)$^{19}\textrm{Ne}$ rate but does not change the bounds on the astrophysically based lower limit set by \cite{Fisker06}. 
This means that the oscillatory burning behavior found in \cite{Fisker06} is most likely not the generator of the ${}^{12}\textrm{C}$ that fuels superbursts \citep{Cumming01}, although such a generator has not yet been found in other multi-zone simulations either \citep{Woosley04,Fisker04,Fisker07a}. 

\subsection{Comparison to observations}
Observations indicate that type I X-ray bursts occur for $\log(I) \lesssim -0.5$ and cease for $\log(I) \gtrsim -0.5$ \citep{Paradijs79,Paradijs88b,Cornelisse03b,Remillard06,Galloway06}.  In contrast with the global linear stability analysis of \cite{Cooper06}, our full network time-dependent simulations show no single trial value of $f$ that is consistent with observations over the entire range of accretion rates we consider. 
This suggests that more complex physics must be included in the simulation. The most important parameters of the bursting behavior as a function of accretion rate are the accreted composition -- here assumed to be solar --  and the mass and radius of the neutron star. 
Other possible factors include the sedimentation of CNO ions \citep{Peng07} as well as  the geometry of the flow from the accretion disk through the boundary layer and onto the neutron star \citep{Inogamov99}. Here the flow of the accreted matter may change as the thermonuclear burning transitions from unstable bursts to stable burning.

\section{The new ${}^{15}\textrm{O}(\alpha,\gamma){}^{19}\textrm{Ne}$ reaction rate}\label{sec:newrate}
The  ${}^{15}\textrm{O}(\alpha,\gamma){}^{19}\textrm{Ne}$ reaction rate is dominated by the resonance contributions at temperatures above $0.1\,\textrm{GK}$  \citep{Langanke86}. 
However, the partial widths $\Gamma_{\alpha}$ and $\Gamma_{\gamma}$ of the
relevant states with excitation energies of 4-5 MeV are not
sufficiently well known for calculating the reaction
rate reliably. Of particular interest is the resonance at an excitation energy of 4.03 MeV in $^{19}$Ne,
just above the $^{15}$O$+\alpha$ threshold. This resonance dominates
the reaction rate at temperatures below 0.6 GK \citep{Langanke86}
and therefore determines the temperature conditions for the
breakout from the hot CNO cycles. Its $\gamma$ width was first measured by \cite{Tan05} and later
confirmed by \cite{Kanungo06}. However, its small $\alpha$-decay
branching ratio of the order $10^{-4}$ \citep[see][for previous upper limits]{Davids03,Rehm03} prevented a reliable estimate of the reaction rate until
the recent work by \cite{Tan07}.

\begin{figure}[tbph]
\center
\includegraphics[width=1.0\linewidth]{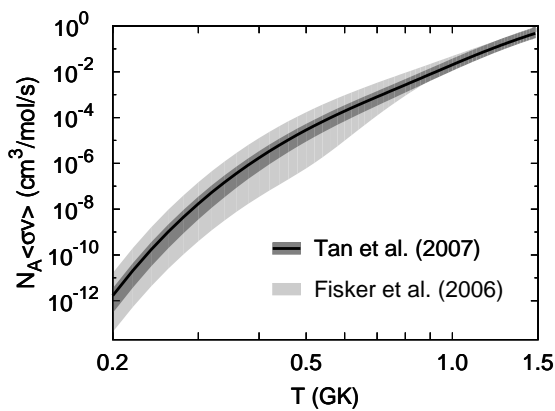}
\caption{\label{fig:rate} Figure shows the newly measured reaction rate along with $1\sigma$ upper and lower experimental uncertainties. The uncertainty range discussed by \cite{Fisker06} is also shown.}
\end{figure}

A new $^{15}$O($\alpha$,$\gamma$)$^{19}$Ne rate was proposed by \cite{Tan07} based on the measured $\alpha$-decay branching ratios and lifetimes of the relevant states in $^{19}$Ne \citep[for details, see][]{Tan07}.
This rate is shown in
Fig.~\ref{fig:rate} with one sigma uncertainty indicated by the darker area, and it is significantly improved as compared
to the previous model-dependent estimate by \cite{Fisker06}. In addition, dominant contributions at higher temperatures from the states at
4.14 and 4.60 MeV were unexpected. This new rate not only allows a better identification of the ignition
conditions of X-ray bursts but also permits the improved analysis
of the dynamics and mechanism of X-ray bursts as demonstrated below.

\subsection{Astrophysical consequences}
\begin{figure}[tbph] \center \includegraphics[width=1.0\linewidth]{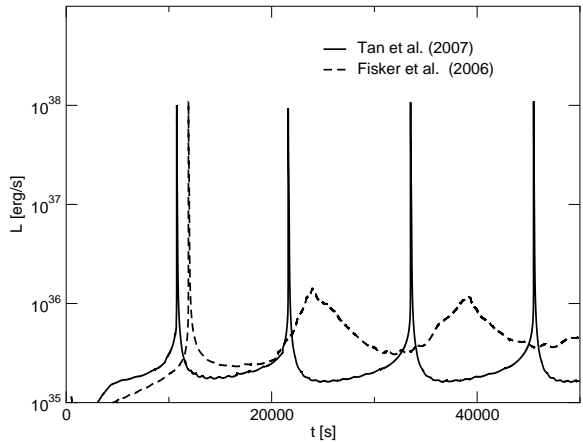}
\caption{\label{fig:nature_fisker06}This figure shows the luminosity as a function of time for a simulation with an accretion rate of $\dot{M}=10^{17} \textrm{g}\,\textrm{s}^{-1}$ comparing the lower limit of the $^{15}$O($\alpha,\gamma$)$^{19}\textrm{Ne}$ for this work vs. the lower limit of the $^{15}$O($\alpha,\gamma$)$^{19}\textrm{Ne}$ rate used by \cite{Fisker06}. We note that the reduced uncertainty of the rate presented in this work corroborates observations and thus constitutes a major improvement of the rate.} 
\end{figure}

Repeating the calculations of \cite{Fisker06} for the lower limit of the $^{15}$O($\alpha,\gamma$)$^{19}\textrm{Ne}$ reaction rate, Fig.~\ref{fig:nature_fisker06} shows a comparison between the luminosity as a function of time for the previous lower limit and the present experimental lower limit for a constant accretion rate. The results show that the new rate within its experimental uncertainties is sufficient to trigger the observed sequences of bursts for a constant accretion rate.

By doing a parameter study covering the previous experimental uncertainty range for the $^{15}$O($\alpha,\gamma$)$^{19}\textrm{Ne}$ reaction it was shown in section~\ref{sec:parameterstudy} and in \cite{Fisker06} that past X-ray burst models have been subject to a very large uncertainty due to the uncertainty of the $^{15}$O($\alpha,\gamma$)$^{19}\textrm{Ne}$ reaction rate.
Fortunately, the widely used rate of \cite{Caughlan88} (viz.~$\log(f)\equiv 0$) is very close to the newly measured rate of \cite{Tan07}. This means that the conclusions of earlier calculations \citep{Fushiki87,Bildsten98c,Fisker03,Woosley04,Heger05,Cooper06} hold.
 
Since the ${}^{15}\textrm{O}(\alpha,\gamma){}^{19}\textrm{Ne}$-reaction rate now has experimentally determined upper as well as lower limits, we can determine the accuracy of the theoretical estimate of the critical transition accretion rate between the steady state burning phase and the burst phase in accreting neutron star binary systems. 

We already noted from Fig.~\ref{fig:constraint} (see subsection \ref{subsec:highfanyI}) that $\log(I)_{crit.}=$ 0.2--0.4.
This corroborates the majority of previous simulations~\citep{Rembges99,Fisker03,Heger05} and calculations~\citep{Fujimoto81,Bildsten98c} which have determined the transition point to be around $\dot{M}\sim 2.1\times 10^{18} \textrm{g}\,\textrm{s}^{-1}$ for a fiducial neutron star with $R=10$ km and $M=1.4M_\odot$ and adopting a solar composition for the accreted matter. 

\begin{figure}[btph]\center\includegraphics[width=1.0\linewidth]{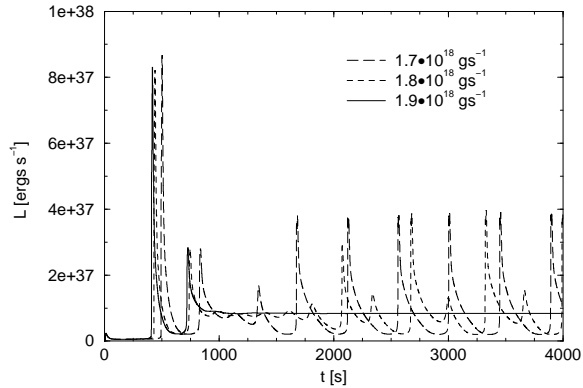}
\caption{\label{fig:wrec_Lt}We used the newly measured rate to calculate the luminosity originating from the nuclear burning as a function of time for different accretion rates. As is seen from the constant luminosity on the graph, the burning is stable for $\dot{M}\geq 1.9\times 10^{18} \textrm{g}\,\textrm{s}^{-1}$. }\end{figure}

Using the rate of \cite{Tan07}, several simulations were run for different accretion rates in the $\log(I)_{crit.}=$0.2--0.4 range while tracking the luminosity resulting from the nuclear burning. 
The results shown in Fig.~\ref{fig:wrec_Lt} show that the burning becomes stable for $\dot{M}\geq 1.9\times 10^{18} \textrm{g}\,\textrm{s}^{-1}$.
Identical simulations were performed using the one sigma upper and lower limits of the newly measured reaction rate as shown in Fig.~\ref{fig:rate}. While the upper limit yields the same transition accretion rate, the lower limit increases the transition point to $\dot{M}\approx 2.1\times 10^{18} \textrm{g}\,\textrm{s}^{-1}$. The astrophysical uncertainty in the determination of the accretion rate at the transition point has thus been reduced to less than 10\% compared to previous rate-induced uncertainties of one order of magnitude in the accretion rate c.f.~Fig.~\ref{fig:constraint} and \cite{Fisker06}.

\section{Conclusion}\label{sec:conclusion}
\noindent
Three important points:
\begin{enumerate}
\item We corroborate the stability analysis of \cite{Cooper06} showing that the atmosphere is stable towards runaways if the ${}^{15}\textrm{O}(\alpha,\gamma)$ ${}^{19}\textrm{Ne}$ reaction rate is low. However, we also show that this instability does not lead to observable bursts if the ${}^{15}\textrm{O}(\alpha,\gamma)$ ${}^{19}\textrm{Ne}$ reaction rate is low.
\item The new measurement is close to the previous and widely used rate of \cite{Caughlan88}, so the conclusions of previous X-ray burst simulations \citep[e.g.][]{Fushiki87,Bildsten98c,Fisker03,Woosley04,Heger05,Fisker07a} do not change.
\item The new measurement of the ${}^{15}\textrm{O}(\alpha,\gamma){}^{19}\textrm{Ne}$ reaction rate significantly reduces the model uncertainty but does not result in a value that is in accord with astronomical observations. Therefore further studies of the other determinants (mass, radius, accretion composition, neutron star core, sedimentation, and possibly the accretion geometry) are needed. Such studies are currently underway.
\end{enumerate}

The simulations in this paper demonstrate why it is important to consider the uncertainty associated with the input parameters of an X-ray burst simulation as it can significantly influence the predicted observables. Furthermore, they show how experimental nuclear data can complement observational results for a better understanding of the complex interplay between the fuel supply and burning processes on the surface of accreting neutron stars.

\begin{acknowledgments}
We would like to thank Anthony L.~Piro for discussions and the referee for helpful suggestions that improved the clarity of the paper. This work is supported by the National Science Foundation under grant No. PHY01-40324 and the Joint Institute for Nuclear Astrophysics\footnote{\emph{www.jinaweb.org}}, NSF-PFC under grant No. PHY02-16783.  R. L. C. is supported in part by the National Science Foundation under Grant No. PHY99-07949.
\end{acknowledgments}


\end{document}